\documentclass[aps, pra, twocolumn, floatfix, superscriptaddress]{revtex4-1}
\usepackage{amssymb, amsmath, bm, graphicx, color, xfrac} %floatrow
\DeclareGraphicsExtensions{.pdf,.png,.jpg}
\newcommand{\gvec}[1]{\boldsymbol{\mathrm{#1}}}
\begin{document}
\title{Collision dynamics of two-dimensional non-Abelian vortices}
\author{Thomas Mawson}
\author{Timothy C. Petersen}
\author{Tapio Simula}
\affiliation{School of Physics and Astronomy, Monash University, Victoria 3800, Australia}
\date{\today}

\begin{abstract}
We study computationally the collision dynamics of vortices in a two-dimensional spin-2 Bose--Einstein condensate. In contrast to Abelian vortex pairs, which annihilate or pass through each other, we observe non-Abelian vortex pairs to undergo \emph{rungihilation}---an event that converts the colliding vortices into a rung vortex. The resulting rung defect subsequently decays to another pair of non-Abelian vortices of \emph{different type}, accompanied by a magnetization reversal.
 
\end{abstract}
\maketitle

% =================================================================
% =================================================================

\section{Introduction}
The quantized vortex is the characteristic topological defect of a Bose--Einstein condensate (BEC) described by a complex scalar order parameter with $U(1)$ phase symmetry \cite{Onsager1949,FEYNMAN195517}. Spinor condensates \cite{Ho1998a, Ohmi1998a} exhibit a variety of order parameter symmetries and correspondingly a greater number of possible topological defects \cite{0305-4470-39-23-017,Kawaguchi2012253}. In spin-1 condensates, topological defects including monopoles \cite{Ray2014a, Ray544,PhysRevX.7.021023,2016arXiv161006228S}, skyrmions \cite{Choi2012a}, vortex knots \cite{hall_spinor_knot} and half-quantum vortices \cite{PhysRevLett.115.015301} have been discovered. In spin-2 BECs the cyclic and biaxial nematic ground states \cite{Ciobanu2000a} permit order parameter symmetries with fractional-charge vortices \cite{semenoff_discrete_2007, PhysRevLett.103.115301, Huhtamaki2009a}. Their non-Abelian algebra opens pathways for studying the reaction dynamics of non-commutative topological defects. Cold atom superfluids provide controllable laboratory systems for studying non-Abelian defects which are also predicted to occur e.g. in models of cosmology \cite{Copeland2007a}, biaxial nematic liquid crystals \cite{PhysRevE.66.051705, PhysRevE.51.1216}, neutron stars \cite{masuda_magnetic_2016} and d-wave Fermi condensates \cite{doi:10.1143/JPSJ.78.113301}. 

The algebra of a pair of vortices manifests in their collision dynamics. The scalar vortices may be described by the phenomenological model introduced by Feynman \cite{FEYNMAN195517}. The collision of such vortex lines induces a reconnection---a topological reaction in which two colliding vortices exchange line sections \cite{PhysRevB.31.5782, PhysRevLett.71.1375,PhysRevLett.86.1410}. Such vortex reconnections have been recently imaged in superfluid helium for the first time \cite{Bewley16092008, Fonda25032014}. In a two-dimensional (2D) superfluid, the reconnection event is replaced by vortex-antivortex annihilation---a process which is well understood in both scalar \cite{PhysRevLett.104.160401, PhysRevA.90.063627,PhysRevA.93.043614} and spin-1 condensates \cite{PhysRevA.94.063615, PhysRevA.93.013615,PhysRevLett.116.185301}. However, the collision dynamics of non-Abelian vortices are not accounted for by the Feynman model. Rather, collisions of certain non-Abelian vortices are topologically constrained to create a rung vortex bridging the two vortices. In three-dimensional (3D) spin-2 BECs, such rung formation remains to be observed experimentally, having been demonstrated numerically for the cyclic phase in the pioneering work by Kobayashi \emph{et al.} \cite{PhysRevLett.103.115301} and more recently for the biaxial nematic phase by Borgh and Ruostekoski \cite{PhysRevLett.117.275302}. In higher spin systems \cite{PhysRevLett.94.160401,PhysRevLett.107.190401,PhysRevA.85.051606} non-Abelian collision dynamics are likely a common occurrence. Vortex collisions are a particularly important mechanism in quantum turbulence \cite{FEYNMAN195517}. The unusual collision dynamics of non-Abelian vortices in 3D with an associated helicity cascade has opened a new area of non-Abelian quantum turbulence \cite{2016arXiv160607190K}. The reaction dynamics of non-Abelian vortices in 2D spin-2 BECs have remained unexplored.

\begin{figure}[t]
\centering
\includegraphics[width=\columnwidth]{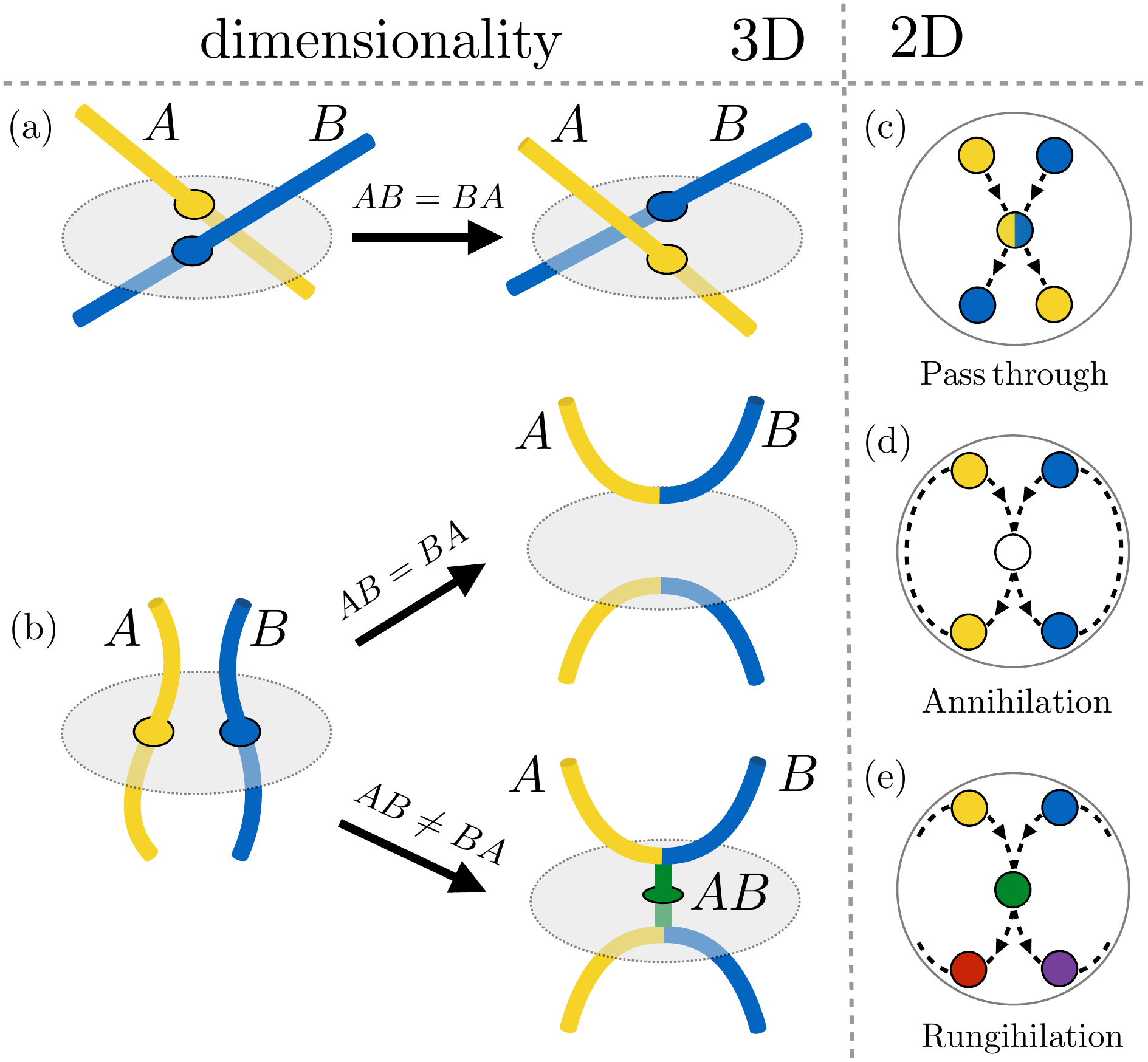}
\caption{ (color online) Schematic of vortex collision dynamics in two and three-dimensional condensates. (a),(b) left: Initial state of vortex lines with invariants $A$ and $B$ piercing a two-dimensional condensate plane, grey shaded disk, that intersects the collision cusp. (a) right: Abelian vortices pass through. (b) right top: Abelian vortices undergo a vortex reconnection. (b) right bottom: Non-Abelian vortices form a rung vortex with invariant $AB$. (c)-(e) Collision of vortices in a two-dimensional condensate, corresponding to the dynamics of the vortex lines on the disk in (a)-(b). Vortices are denoted by circles with color representing the invariant. (c) Passing through, split circle denotes overlapping defects. (d) Annihilation, white circle denotes absence of defects. (e) Rungihilation, green circle denotes the rung defect.}
\label{schem}
\end{figure}

Here we study the collision dynamics of vortices in two-dimensional cyclic spin-2 Bose--Einstein condensates. We consider topological reactions of vortices with either commuting or non-commuting topological invariants, as illustrated in the schematic Fig.~\ref{schem}. As in the experiments by the groups of Anderson \cite{PhysRevLett.104.160401} and Hall \cite{Freilich1182}, the mutual induction field of the vortex pair propels the defects inducing a collision event if the paths overlap. For Abelian vortex pairs we observe vortex-antivortex annihilation or pass through events, depending on the structure of the order parameter. For non-Abelian vortices we identify a new collision event, coined rungihilation, which is a two-dimensional counterpart to rung formation for non-Abelian vortex lines in 3D condensates. Followed by the rungihilation, we observe the rung defect to decay into another pair of non-Abelian vortices of a \emph{different type}. Associated with such vortex mutation we observe reversal of the magnetization of the vortex cores. Rungihilation has potentially interesting consequences for 2D non-Abelian quantum turbulence \cite{mawson_route_2015}. 
 
% =================================================================
% =================================================================

\section{Non-Abelian quantum vortices} 
Within the mean-field theory, the spin-2 BEC order parameter is represented by a spinor wave function $\Psi(\gvec{r},t)$ with five components $\Psi_m$ and total particle density $n\left(\gvec{r},\,t\right) = \sum_{m=-2}^2\left|\Psi_m\right|^2$, where the subscript $m$ denotes the magnetic quantum number. The Hamiltonian density is \cite{Kawaguchi2012253} 
\begin{equation}
\mathcal{H} = K + \frac{c_0}{2}n\left(\gvec{r}\right)^2 + \frac{c_1}{2} \left|\gvec{F}\left(\gvec{r}\right)\right|^2  + \frac{c_2}{2} \left|A\left(\gvec{r}\right)\right|^2,
\end{equation}
where the single particle term $K = \sum_{m=-2}^{2}\Psi_m^*\left[-\hbar^2\nabla^2 / 2M + V_{\rm ext}\right]\Psi_m$ includes an external potential $V_{\rm ext}$ in addition to the kinetic energy term. The s-wave particle interactions with strengths $c_i$ depend on the total density $n\left(\gvec{r}\right)$, the spin density vector $\gvec{F}\left(\gvec{r}\right)$ with components $F_\nu = \sum_{i,j=-2}^{2}\Psi^*_i(f_\nu)_{ij}\Psi_j$, where $f_\nu$ are the spin-2 Pauli matrices, and the spin-singlet pair amplitude $A\left(\gvec{r}\right) = \left(2\Psi_2\Psi_{-2} - 2\Psi_1\Psi_{-1} + \Psi_0^2\right)/\sqrt{5}$. The latter two interactions are spin-dependent. For small external magnetic fields the coupling constants $c_1$ and $c_2$ determine the ground state phase diagram \cite{Ciobanu2000a}. The cyclic phase is obtained for $c_1 > 0$ and $c_2 > 0$, where $|\gvec{F}\left(\gvec{r}\right)| = 0$ and $|A\left(\gvec{r}\right)| = 0$. 

A representative cyclic phase order parameter is given by $\Psi_{\rm{cyclic}} = (i,\,0,\,\sqrt{2},\,0,\,i)^\mathsf{T}/2$ up to a transformation $R = e^{i\phi}e^{-i\gvec{f}\cdot\gvec{\hat{\omega}}\Theta}$ involving a gauge angle $\phi$ and a spin rotation angle $\Theta$ about the $\gvec{\hat{\omega}}$ axis. Group theoretically, the order parameter manifold of the cyclic phase is $G/H = U(1)\times SU(2)/T^*$ where $T^*$ is the non-Abelian binary tetrahedral symmetry group. Topological defects can be characterized by a topological invariant, a transformation that leaves the defect order parameter unchanged. The elements of $T^*$ specify the different vortex types in the cyclic phase and their topological invariants \cite{semenoff_discrete_2007}, up to a relabelling. The invariants, classified in the conjugacy classes (I)-(VII), are: 
\begin{align}
\begin{split}
(\rm{I}) & \, \{(\eta,\,\bf{1})\}, \\ 
(\rm{II}) & \, \{(\eta,\,\bf{-1})\}, \\
(\rm{III}) & \, \{(\eta,\,i\sigma_\nu),\,(\eta,\,-i\sigma_\nu)\}, \\
(\rm{IV}) & \, \{(\eta+1/3,\,\tilde{\sigma} ),\,(\eta+1/3,\,-i\sigma_{\nu}\tilde{\sigma})\}, \\
(\rm{V})  & \, \{(\eta+1/3,\,-\tilde{\sigma} ),\,(\eta+1/3,\,i\sigma_{\nu}\tilde{\sigma} )\}, \\
(\rm{VI}) & \, \{(\eta+2/3,\,-\tilde{\sigma} ^2),\,(\eta+2/3,\,-i\sigma_{\nu}\tilde{\sigma} ^2)\},\\
(\rm{VII}) & \, \{(\eta+2/3,\,\tilde{\sigma} ^2),\,(\eta+2/3,\,i\sigma_{\nu}\tilde{\sigma} ^2)\},
\end{split}
\end{align}
where $\nu = x,\,y,\,z$ and the $SU(2)$ components of the elements of $T^*$ are given by the spin-1/2 Pauli matrices $\sigma_{\nu}$ and $\tilde{\sigma} \equiv (\textbf{1}+\sum_{\nu} i\sigma_{\nu})/2$. The integer $\eta$ is a $U(1)$ rotation winding number. We refer to each vortex by its invariant using the shorthand $\pm$X$_{\eta}^\nu \equiv (\eta + a_X,\, g_{\nu}^X)$, where X is the class number, $a_X$ is a class specific constant and $g^X_\nu= \mathbf{1}\cos(\Theta_X/2)+i(\gvec{\hat{\omega}}\cdot\gvec{\sigma})\sin(\Theta_X/2)$ the $SU(2)$ component of an element of $T^*$. The sign of X$_{\eta}^\nu$ denotes the direction of the spin circulation of $g^X_\nu$. For example $-$VI$_{-1}^{x}\equiv\,(-1+2/3,-i\sigma_x\tilde{\sigma} ^2)$. The invariants can be mapped onto the R transformations as $R(X^\nu_\eta) = \,e^{i2\pi(\eta+a_X)}\,e^{-i\gvec{f}\cdot\gvec{\hat{\omega}}\Theta_X}$. When $\eta = 0$, class (I) represents the vortex free state, while  (II) and (III) are spin vortices with $1$ and $1/2$ a unit of spin circulation, respectively.  Classes (IV),(V) and (VI),(VII) describe a fractional-charge $1/3$ and a $2/3$ vortex, respectively. Topological invariants within a specific class can be transformed into each other via gauge and spin rotations. The vortices can be traced by their core structures, which are either $|A| \neq 0$, classes (II)-(III), or $|F_z| \neq 0$, classes (IV)-(VII), where the magnetization density $F_z = 2(\Psi_2-\Psi_{-2}) + \Psi_1 - \Psi_{-1}$.  

% FIGURE
% FIGURE

% FIGURE
\begin{figure*}[htb]
 \centering
 \includegraphics[width=2\columnwidth]{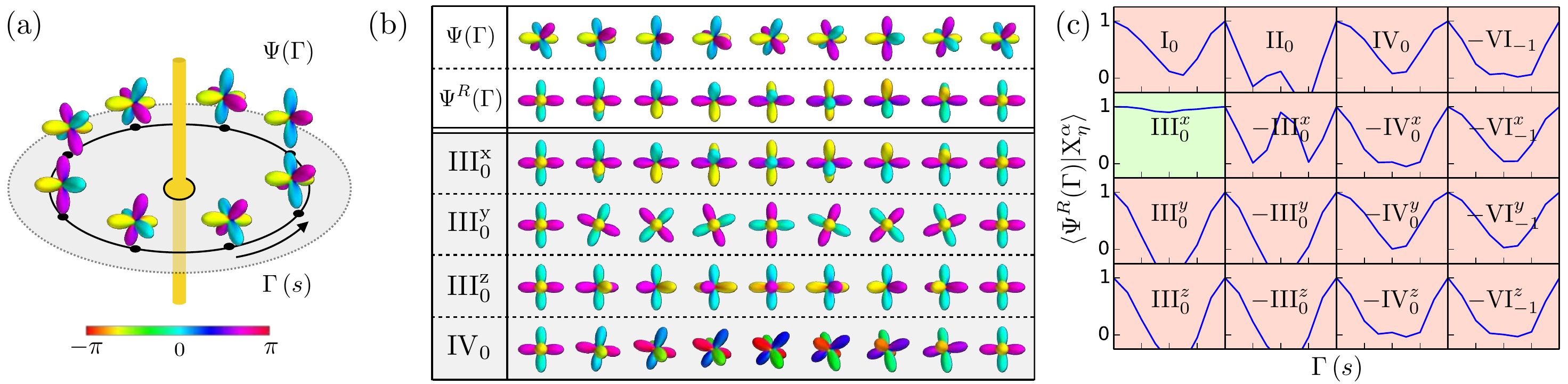}
 \caption{(color online) A method to identify the topological invariant of a vortex. (a) The measured spherical harmonics representation $\Psi\left(\Gamma\right)$ of the vortex to be identified. (b) $\Psi\left(\Gamma\right)$ and its rotated form $\Psi^R\left(\Gamma\right) = e^{i\phi}e^{-i\alpha f_z}e^{-i\beta f_y}e^{-i\gamma f_z}\Psi\left(\Gamma\right)$ for phase angle $\phi=9\pi/7$ and spin angles $\alpha=\pi/2,\,\beta=3\pi/10$ and $\gamma=9\pi/7$. The IV$_0$, and III$_0^{x,y,z}$ vortices of the standard basis are shown for comparison. (c) Overlap integral $\langle\Psi^R\left(\Gamma\right)|\mathrm{X}^\nu_\eta\rangle$ for vortices $\mathrm{X}^\nu_\eta$ of the standard basis. The correctly identified invariant is shaded green.}
 \label{meas}
\end{figure*}
% FIGURE

% =================================================================
% =================================================================

\section{Vortex collision dynamics}
The algebra of the topological invariants for a pair of vortices in the cyclic phase has important implications for their collision dynamics. In Fig.~\ref{schem}(a) and (b) we outline the possible outcomes for collisions of different vortex pairs. When the algebra is Abelian the collision dynamics are similar to those of scalar vortex lines. The usual outcome is a reconnection event where the two vortices join at a cusp, exchange line ends then separate, see the upward arrow in Fig.~\ref{schem}(b). Certain Abelian vortices in spinor condensates exhibit a pass through collision event, where the vortex lines cross apparently without interaction, see Fig.~\ref{schem}(a). For vortices with a non-Abelian algebra, topological constraints enforce rung formation---an event in which the vortex lines remain joined, forming an additional rung vortex. Under unusual energetic circumstances, a rung might also form during a collision of Abelian vortices. The downwards arrow in Figure~\ref{schem}(b) illustrates a collision of non-Abelian vortices with invariants $A$ and $B$, which may result in either a rung vortex with invariant $AB$ (shown) or a $BA^{-1}$ rung, with composition rule $AB = (\eta_A + a_A+\eta_B + a_B, \,g^Ag^B)$. 

In thin condensates the vortex lines behave as two-dimensional point-like defects and their collision dynamics change accordingly. As shown in Fig.~\ref{schem} the dynamics of vortices in 2D are expected to correspond to the motion of the intersection sites between virtual 3D vortex lines and a 2D plane. In 3D systems two initially parallel vortex lines can locally change their relative orientation to initiate a topological reaction such as a reconnection; there exists no such freedom for point-like defects. Therefore, in 2D, only AB type topological reactions of vortices A and B may occur with BA$^{-1}$ events being suppressed by dimensionality and topological invariant conservation. As shown in Fig.~\ref{schem}(c), pass through remains unchanged for point-like defects. Vortex reconnection becomes vortex-antivortex annihilation where, as shown in Fig.~\ref{schem}(d), the collision leaves the condensate defect free. For non-Abelian vortex pairs we anticipate a new collision event, coined rungihilation. As shown in Fig.~\ref{schem}(e), rungihilation is a 2D equivalent of rung formation dynamics of 3D vortices in which two non-Abelian defects collide and fuse into a non-trivial rung defect with invariant $AB$.
 
% =================================================================
% =================================================================

\section{Vortex identification}
The topological invariant of a vortex can be deduced by considering a closed loop $\Gamma\left(s\right)$ with curve parameter $s$ around the vortex core in real space, which is mapped to a closed loop $\Psi(\Gamma)$ in the order parameter space \cite{mermin_topological_1979}. Loops that can be smoothly deformed into one another are homotopic and form the elements of the first homotopy group. The topological invariant is defined as a composite gauge and spin rotation $R$, which preserves the single valuedness of the vortex order parameter at the arbitrarily chosen base point of the closed loop. 

For a single vortex the topological invariant can only be defined up to its class. For systems with two or more vortices we choose loops, separately encircling each vortex, which share the same fixed base point. We assign topological invariants to vortices in our simulations by representing the vortex order parameter in terms of a spherical harmonic decomposition, $\Psi(\Gamma) = \sum_{m=-2}^2 \Psi_m(\Gamma) Y^m_2(\gvec{r})$ \cite{Kawaguchi2012253}. The result is a series of geometric objects that reveals the identity of the vortex and may be used for visualising the characteristic gauge and spin rotations of the vortex, see Fig.~\ref{meas}. At the base point we choose a non-unique rotation $R$, which transforms the measured series $\Psi(\Gamma)$ to one series $\Psi^R(\Gamma)$ of a standard basis, provided as Fig.~S1 in the Supplementary Materials \cite{supplemental}. The vortex is identified by comparing the transformed series $\Psi^R(\Gamma)$ with each of the series $\mathrm{X}^\nu_\eta$ in the standard basis. The comparison takes the form of an overlap integral $\langle\Psi^R\left(\Gamma\right)|\mathrm{X}^\nu_\eta\rangle$ performed at each point $\Gamma(s)$ on the loop, see Fig.~\ref{meas}(c). 

% =================================================================
% =================================================================

\section{Numerical experiments}
To investigate the vortex pair collision dynamics we use XMDS2 \cite{dennis2013xmds2} to numerically solve the 2D spin-2 Gross--Pitaevskii equation \cite{Kawaguchi2012253} for a condensate of $^{87}$Rb atoms with particle number $N = 75000$ on a mesh with $2048\times2048$ grid points. The condensate is trapped in a harmonic potential with trap frequency $\omega_{\rm trap} = 2\pi \times 200$ Hz. Unless otherwise stated the numerical results are presented in terms of dimensionless quantities with units of time $\tau=1/\omega_{\rm trap}$ and space $l=\sqrt{\hbar/2M\omega_{\rm trap}}$, where $M$ is the atomic mass. The dimensionless coupling constant $\tilde{c}_0 = c_0N/\hbar\omega_{\rm trap}l^2=0.231N$, where $c_0=4\pi \hbar^2(4a_2+3a_4)/7M$ is specified by the experimentally measured scattering lengths $a_i$ of $^{87}$Rb \cite{PhysRevA.64.053602}. Following Kobayashi \emph{et al.} \cite{PhysRevLett.103.115301} we choose $c_1 = c_2 = 0.5\, c_0$ in comparison to the typical $^{87}$Rb values of $c_2 \simeq 0.0103\,c_0$ and $c_2 \simeq -0.0055\,c_0$ \cite{1367-2630-8-8-152}. Despite the large values of $c_1$ and $c_2$ which presently pose a challenge to experimentalists, these coupling constants are a theoretically justified choice to ensure that the condensate is deep in the cyclic phase, hence isolating topological effects of the vortex dynamics from their energetics. In the following we describe the simulation results for vortex-antivortex annihilation, pass through, and rungihilation for the given representative vortex pairs: the Abelian pair collisions;  [IV$_0$, $-$VI$_{-1}$], and [IV$_0$, IV$_{-1}$]; and a non-Abelian pair collision [IV$_0$, $-$VI$_{-1}^{y}$]. The simulations are presented in the Supplementary Materials \cite{supplemental}. 

% =================================================================
% =================================================================

% FIGURE
\begin{figure*}[ht]
 \centering
 \includegraphics[width=2\columnwidth]{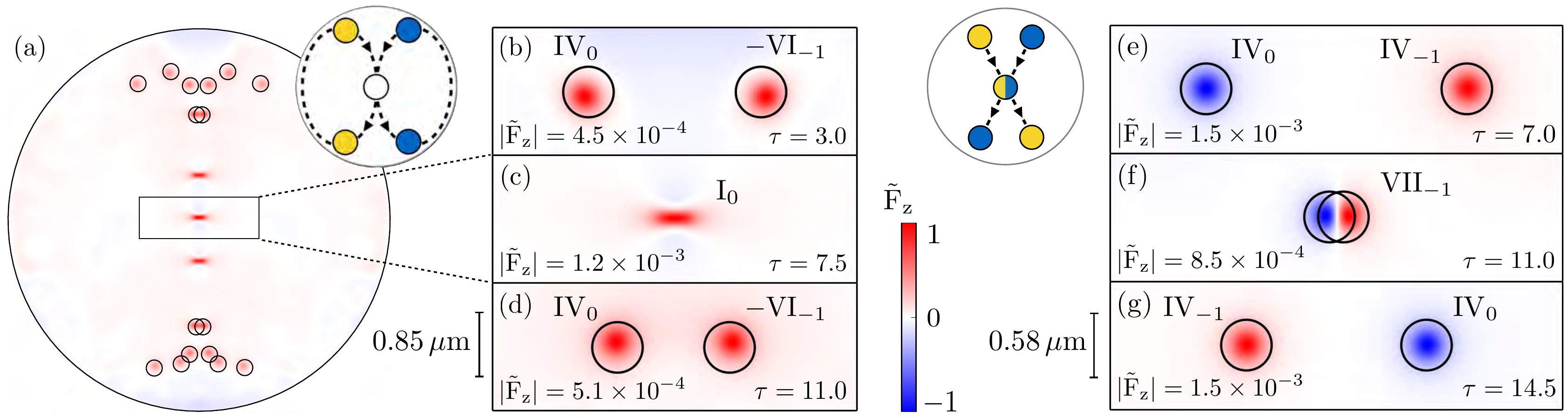}
 \caption{(color online) Collision dynamics of vortices with Abelian algebra. (a)-(d) Annihilation of IV$_0$ and $-$VI$_{-1}$ vortices. (a) Motion of the vortices through the condensate traced by their magnetized cores. The figure is a composite created by overlaying the normalized sum of $\tilde{F}_z$ at eleven different times. (b)-(d) Vortices at times $\tau$ prior, during and post the collision event. (e)-(g) Pass through of IV$_0$ and IV$_{-1}$ vortices. The locations of the vortices are represented by the circles surrounding the magnetic vortex cores and the normalization constant of the magnetization is provided in each frame (b)-(g).}
 \label{fig3}
\end{figure*}
% FIGURE

\subsection{Annihilation}
We consider vortex-antivortex annihilation for an Abelian vortex pair with invariants IV$_0$ and $-$VI$_{-1}$. The representative order parameters are $\Psi(\gvec{r};\mathrm{IV}_0) = e^{i\theta/3}\,e^{-i\theta (f_x + f_y + f_z)/3\sqrt{3}}\Psi_{\rm{cyclic}}$ and $\Psi(\gvec{r};\mathrm{-VI}_{-1})  = e^{-i\theta/3}\,e^{i\theta (f_x + f_y + f_z )/3\sqrt{3}}\Psi_{\rm{cyclic}}$, where $\theta$ is the polar angle. Each vortex has a magnetized core with $F_z > 0$, which is useful for tracing the vortex paths shown in Fig.~\ref{fig3}. The vortices, driven by their mutual induction field, travel along paths which overlap at $\tau = 6.0$ and undergo vortex-antivortex annihilation. The annihilation is survived by a remnant magnetic vortexonium highlighted in Fig.~\ref{fig3}(c). The magnetic vortexonium is a spatially localized bound state of an Abelian vortex-antivortex pair and is a generalization of the Jones-Roberts soliton of scalar BECs \cite{PhysRevA.93.043614,PhysRevA.90.063627,PhysRevA.91.013612, 0305-4470-15-8-036}. Topologically, the magnetic vortexonium is equivalent to the trivial defect I$_0$. The vortex pair is reformed when the vortexonium travels into the low density boundary region of the condensate where vortex pair creation becomes energetically feasible. We have confirmed, by measurement, that the reformed vortices have invariants IV$_0$ and $-$VI$_{-1}$.

% =================================================================
% =================================================================

\subsection{Pass through} 
While pass through is topologically permissible for Abelian vortices, its occurrence depends on the vortex kinematic details. For example, for scalar vortex-vortex (antivortex-antivortex) pairs pass through is hindered by the Coulomb-like repulsive interaction arising from the energy barrier associated with the superflow mass currents. In spinor BECs the superflow mass currents of the two vortices may be associated with different spin-components. Consequently, the repulsive energy barrier may not exist thereby allowing the vortices to pass through. Figure~\ref{fig3}(e)-(g) shows a pass through event for two Abelian vortices with invariants IV$_0$ and IV$_{-1}$. To generate such vortex wave functions we consider a second cyclic ground state $\Psi' = e^{-i {\cos^{-1}}(1/\sqrt{3}) f_y}\,e^{-i\pi f_z/4}\,\Psi_{\rm{cyclic}}$. By applying appropriate gauge and spin rotations to $\Psi'$ we obtain vortex wave functions $\Psi(\gvec{r};\mathrm{IV}_0)  = e^{i\theta/3}\,e^{i\theta f_z /3}\Psi' $ and $\Psi(\gvec{r};\mathrm{IV}_{-1}) = e^{-i2\theta/3}\,e^{i\theta f_z /3} \Psi'$ with magnetized cores of $F_z = -\sqrt{2/3}$ and $F_z = 2/\sqrt{3}$, respectively. The vortices are driven towards each other by the interactions with their respective image vortices. The vortices overlap at $\tau = 11.0$ and then pass through each other. During the overlap the topology is identified by the total invariant given by IV$_0$IV$_{-1}=$ VII$_{-1}$. A clean pass through was observed in simulations with initial states having identically zero population in the unpopulated components of $\Psi(\gvec{r};\mathrm{IV}_{0})$ and $\Psi(\gvec{r};\mathrm{IV}_{-1})$. 

% =================================================================
% =================================================================

% FIGURE
\begin{figure}[t]
 \centering
 \includegraphics[width=\columnwidth]{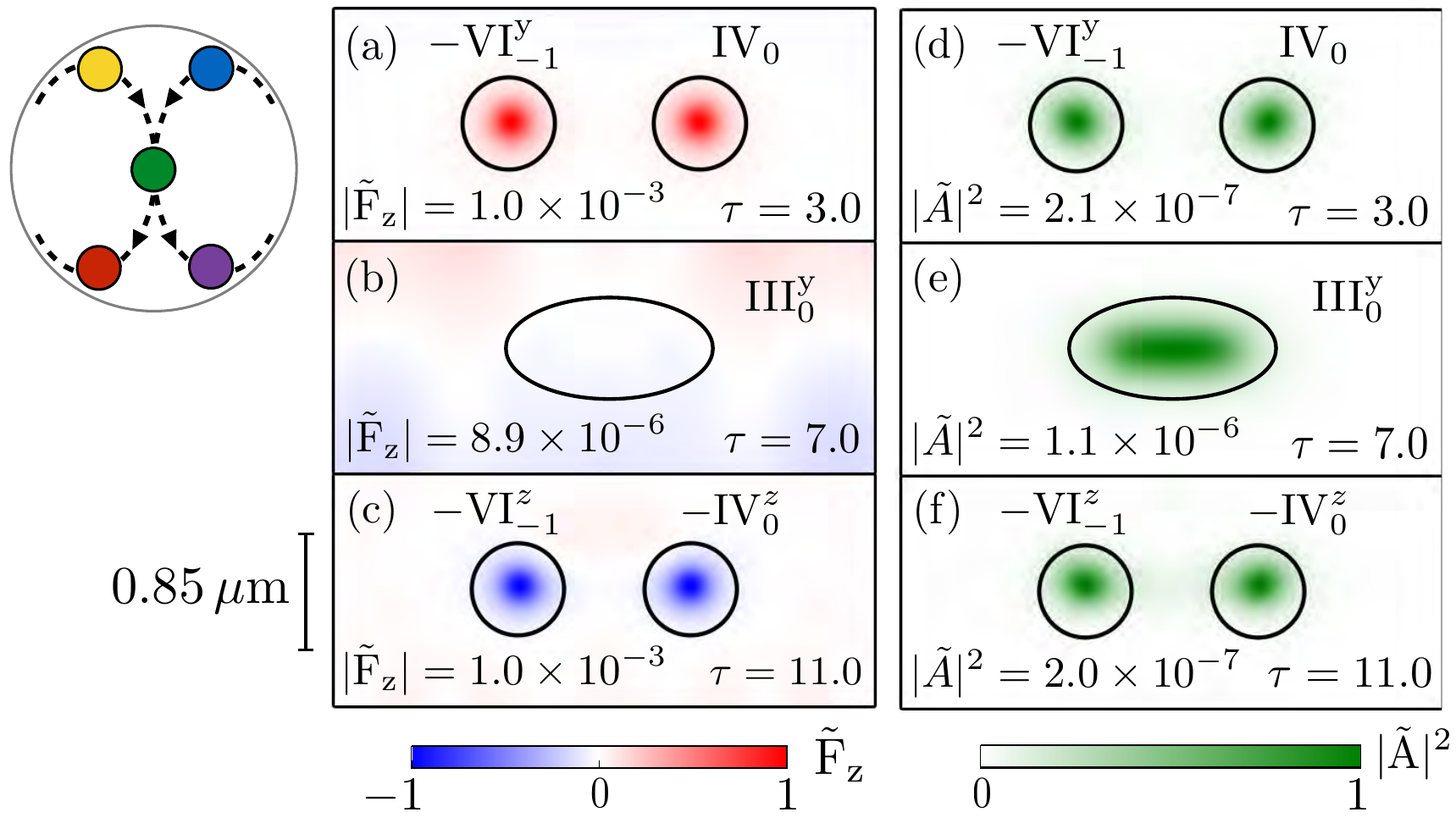}
 \caption{(color online) Rungihilation of IV$_0$ and $-$VI$^y_{-1}$ vortices with non-Abelian algebra. (a)-(f) Vortices at times $\tau$ prior, during and post the collision event. (a)-(c) The normalized magnetization density $\tilde{F}_z$. (d)-(f) The normalized spin-singlet amplitude $|\tilde{A}|^2$. The locations of the vortices are represented by the circles, the rung is highlighted by the ellipses, and the normalization constants $|\tilde{F}_z|$ and $|\tilde{A}|^2$ are provided for each frame.}
  \label{fig5}
\end{figure}
% FIGURE

\subsection{Rungihilation}
The non-Abelian vortex pair  IV$_0$ and $-$VI$^y_{-1}$ are initialised with order parameters $\Psi(\gvec{r};\mathrm{IV}_0)  = e^{i\theta/3}\,e^{-i\theta (f_x + f_y + f_z )/3\sqrt{3}}\Psi_{\rm{cyclic}}$ and $\Psi(\gvec{r};\mathrm{-VI}^y_{-1}) = e^{-i\theta/3}\,e^{i\theta (-f_x - f_y + f_z )/3\sqrt{3}}\Psi_{\rm{cyclic}}$, respectively. Both vortices have a core structure with $F_z > 0$ as shown in Fig.~\ref{fig5}(a). The vortex pair collides forming a rung defect with $F_z=0$ and non-zero $|A|$ core highlighted in Figs.~\ref{fig5}(b) and (e). The rung has a topological invariant $-$VI$^y_{-1}$IV$_0 = $ III$^y_0$. Since an isolated rung defect can only be classified up to its class, a particular topological invariant is ascribed according to the result of $AB$, where $A$ and $B$ are the measured invariants of the vortices before their collision. The rung exists for $\tau=5.0$ time units before breaking up into a pair of \emph{different type} of non-Abelian vortices. The non-Abelian rungihilation and subsequent pair-creation process not only changes the vortex types but is also accompanied by the reversal of the direction of the magnetized core structure, cf. Figs.~\ref{fig5}(a) and (c), such that the new vortices have cores with $F_z < 0$, providing an experimentally detectable signal of this unconventional topological reaction. The newly spawned vortices are measured to have invariants $-$VI$^z_{-1}$ and $-$IV$^z_0$ with total invariant $(-$VI$^z_{-1})$$(-$IV$^z_0) = $ III$^y_0$.

% =================================================================
% =================================================================

\section{Conclusions} 
We have investigated the collision dynamics of vortex pairs in the cyclic state of two-dimensional spin-2 Bose--Einstein condensates. We have shown using numerical experiments that the collision of a pair of vortices may result either in annihilation, passing through, or rungihilation of the two vortices. We have identified the rungihilation event as the two-dimensional counterpart to rung formation of three-dimensional non-Abelian vortices \cite{PhysRevLett.103.115301,PhysRevLett.117.275302}. The non-Abelian vortex pairs could potentially be prepared experimentally using methods similar to those outlined in \cite{PhysRevLett.117.275302}. Imaging the spin-singlet pair amplitude, useful for characterising the rung defects, remains an experimental challenge. However, the rungihilation could possibly be inferred by the associated reversal of the vortex core magnetization using magnetization sensitive imaging techniques \cite{PhysRevLett.95.050401}. Rung formation has opened a new research area of three-dimensional non-Abelian quantum turbulence typified by a novel helicity cascade \cite{2016arXiv160607190K}. The rungihilation of two-dimensional non-Abelian vortices is anticipated to have interesting ramifications for energy flow in 2D quantum turbulent states involving non-Abelian vortices.

% =================================================================
% =================================================================

\begin{acknowledgments}
We are grateful to Michikazu Kobayashi and Gary Ruben for useful discussions and technical support. We acknowledge financial support from the Research Training Program (T.M.) and the Australian Research Council via Discovery Projects No. DP130102321 and No. DP170104180 (T.S.).
\end{acknowledgments}

\end{document}